\documentclass[11pt,epsf,fleqn]{article}
\usepackage{epsfig}
\usepackage[dvips]{color}
\begin{document}
\title{Electroweak baryogenesis in the MSSM with vector-like quarks}
\author{S. W. Ham$^{(1)}$\footnote{s.w.ham@hotmail.com},
Seong-A Shim$^{(2)}$\footnote{shims@sungshin.ac.kr},
and S. K. Oh$^{(3)}$\footnote{sunkun@konkuk.ac.kr}
\\
\\
{\it (1) School of Physics, KIAS, Seoul 130-722, Korea} \\
{\it (2) Department of Mathematics, Sungshin Women's University} \\
{\it Seoul 136-742, Korea} \\
{\it (3) Department of Physics, Konkuk University, Seoul 143-701, Korea}
\\
\\
}
\date{}
\maketitle
\begin{abstract}
In the minimal supersymmetric standard model (MSSM), a strongly first-order electroweak phase transition (EWPT)
is only possible in a confined parameter region where one of the scalar top quarks is lighter than the top quark
and the other one is as heavy as the SUSY breaking scale.
If the MSSM is enlarged to accommodate vector-like quarks and their superpartners, we find that
the strongly first-order EWPT is possible
without requiring light scalar top quark at the one-loop level,
in the limit where the lightest scalar Higgs boson of the MSSM behaves like
the Higgs boson of the standard model
and the other Higgs bosons are all as heavy as the SUSY breaking scale.
The strength of the first-order EWPT is found to be dependent on the mass of the lightest
neutral Higgs boson and the mixing effects of the vector-like scalar quarks.
\end{abstract}
\vfil\eject

\section{Introduction}
Sakharov has suggested several decades ago that the baryon asymmetry of the universe may
be dynamically generated if three conditions are met [1].
The three Sakharov conditions are the violation of the baryon number conservation,
the violation of both C and CP, and the deviation from thermal equilibrium.
Among them, the deviation from thermal equilibrium is the most intricate and challenging condition
for theoretical models to satisfy.
In order to ensure sufficient deviation from thermal equilibrium,
the electroweak phase transition (EWPT) should be strongly first order, since otherwise
the baryon asymmetry generated
during the EWPT would subsequently disappear.

In general, the electroweak symmetry of the universe is spontaneously broken as the universe is
cooled down from very high temperature to the present temperature.
In between the cooling process, the shape of the potential of the universe
at a critical temperature would have two local minima for the first-order EWPT.
One minimum corresponds to the original symmetric-phase state and the other one the broken-phase state.
The strength of the EWPT is measured by comparing the vacuum expectation value (VEV)
of the Higgs field at the broken-phase state with the critical temperature.
If the critical VEV is larger than the critical temperature, the first-order EWPT is said to be strong.

A number of supersymmetric models have been investigated within the context of the EWPT for
the electroweak baryogenesis.
Several authors have investigated the MSSM to show that a strongly first-order EWPT is only
possible in a confined parameter region where a scalar top quark is lighter than the top quark and
the other scalar top quark is as heavy as the SUSY breaking scale of 2 TeV [2-5].
Then, in order to avoid the light scalar top quark, other models extended from the MSSM have also been
investigated for the possibility of a strongly first-order EWPT,
such as the non-MSSMs [6-19] and the MSSM with four generations of fermions [20,21].

The non-MSSMs are different from the MSSM in the sense that there is a trilinear term of the singlet Higgs field
accounting for the effective $\mu$ term.
The trilinear term between Higgs singlet and doublets of the non-MSSMs allows
the mass of the scalar top quark to be sufficiently large.
Thus, the non-MSSMs have some parameter regions where a sufficiently strong first-order EWPT
may occur for electroweak baryogenesis, without requiring a light scalar top quark.
Recently, it has been found that if the fourth generation of fermions is introduced to the MSSM,
it is possible to have a strongly first-order EWPT without a light scalar top quark in
terms of the thermal contribution of the scalar quarks of the fourth generation [20,21].

In this article, we study another model, the MSSM with vector-like quarks
within the context of the electroweak baryogenesis.
It has been proposed some years ago [22,23], and its Higgs sector has been
extensively investigated recently [24-28].
There are three ways to introduce the chiral superfields to the MSSM, maintaining the perturbative
gauge unification with the masses of new extra chiral supermultiplets at the TeV scale [28].
Among them, we pay attention to the specific model, where two vector-like quarks are
present.
This model is similar to the QUE model in Ref. [28], except that we do not consider
an additional chiral lepton superfield.

Our model is therefore actually the MSSM with two additional chiral superfields,
${\cal Q}$ and ${\cal U}$, which are defined to transform under
$SU(3)_C \times SU(2)_L \times U(1)_Y$ as
\begin{eqnarray}
& & {\cal Q} = ({\bf 3}, {\bf 2}, 1/6),  \quad {\bar {\cal Q}} = ({\bar {\bf 3}}, {\bf 2}, - 1/6) \ , \cr
& & {\cal U} = ({\bf 3}, {\bf 1}, 2/3),  \quad {\bar {\cal U}} = ({\bar {\bf 3}}, {\bf 1}, - 2/3) \ .
\end{eqnarray}
These additional chiral superfields introduce new physical vector-like quarks, namely,
two charge $+ 2/3$ quarks $T_1$, $T_2$, and one charge $- 1/3$ quark $B$,
together with their scalar superpartners, ${\tilde T}_i$ ($i = 1,2,3,4$) and ${\tilde B}$ ($i = 1,2$).

The presence of the vector-like sector in the MSSM alters the phenomenology of the MSSM
with respect to the EWPT.
Our study shows that in our model a strongly first-order EWPT is possible at the one-loop level,
without requiring a light scalar top quark.

\section{Higgs potential}

The superpotential of our model may be written as
\begin{eqnarray}
    {\cal W} & = & {\cal W}_{\rm MSSM} + M_Q {\cal Q} {\bar {\cal Q}}
    + M_U {\cal U} {\bar {\cal U}} + Y_u {\cal H}_u {\cal Q} {\bar {\cal U}}
    - Y_d {\cal H}_d {\bar {\cal Q}} {\cal U} \ ,
\end{eqnarray}
where ${\cal W}_{\rm MSSM}$ is the superpotential of the MSSM,
$M_Q$ and $M_U$ are the masses of the vector-like quarks,
and $Y_u$ and $Y_d$ are the Yukawa coupling coefficients
to the weak hypercharge $- 1/2$ and $1/2$ Higgs doublets
$H_d^T = (H_d^0, H_d^-)$ and $H_u^T = (H_u^+, H_u^0)$
in the Higgs superfields ${\cal H}_u$ and ${\cal H}_d$, respectively.

The tree-level Higgs potential of the MSSM is given as
\begin{eqnarray}
V^0 & = & m_u^2 |H_u|^2 + m_d^2 |H_d|^2 - \left (m_{ud}^2 H_u H_d + {\rm H.c.} \right )
+ {1 \over 8} ({g_1}^2 + g_2^2) \left ( |H_u|^2 - |H_d|^2  \right )^2    \  ,
\end{eqnarray}
where $m_u$, $m_d$, and $m_{ud}$ are the mass parameters,
$g_1$ and $g_2$ are the gauge coupling coefficients for $U(1)$ and $SU(2)$, respectively.
Two of the mass parameters, $m_d$ and $m_u$, may be eliminated using the minimum equations
with respect to the neutral Higgs fields.

After electroweak symmetry breaking, as the neutral Higgs fields develop VEVs,
$v_d = \langle H_d \rangle$ and $v_u = \langle H_u \rangle$,
five physical Higgs bosons emerge:
two neutral scalar Higgs bosons, $h$ and $H$ ($m_h < m_H$),
one neutral pseudoscalar Higgs boson, $A$,
and a pair of charged Higgs bosons, $H^{\pm}$.
We assume that the VEVs as well as the parameters of the Higgs potential are all real.
Note that the tree-level mass of $h$ can be expressed in terms of just two parameters,
$m_{ud}$ and $\tan \beta$, where $\tan \beta  = v_u/v_d$.
The tree-level mass of $h$ is smaller than the $Z$ boson mass.

Let us consider the decoupling limit where $h$ remains light at the electroweak scale and all the other
Higgs bosons become very heavy with masses of the order of the SUSY breaking scale or a few TeV.
In this limit, the Higgs sector reduces to the Higgs sector of the Standard Model (SM)
at the electroweak scale and the coupling coefficients of $h$ to gauge bosons and fermions are identical
to the couplings of the SM Higgs boson, $\phi$.
Thus, in the decoupling limit, one cannot distinguish the phenomenology of $h$ from that of $\phi$, hence
the same experimental lower bound on the mass of $h$ as the SM Higgs boson
which is about 114.4 GeV as set by the LEP2 experiments [29].

Hereafter, we confine ourselves in the decoupling limit and denote
the lighter neutral scalar Higgs boson as $\phi$:
\begin{equation}
\phi = \cos \beta {\rm Re} (H_u^0) + \sin \beta {\rm Re} (H_d^0) \ .
\end{equation}
In terms of $\phi$, the tree-level Higgs potential at zero temperature may be written as
\begin{equation}
    V_0(\phi, 0) =  - {m^2_0 \over 2} \phi^2 + {\lambda \over 4} \phi^4 \ ,
\end{equation}
where $m_0$ is related to the mass of the lighter neutral scalar Higgs boson at the tree level
and $\lambda = (g_1^2 + g_2^2) \cos^2 2 \beta/4$ represents
the quartic Higgs self-coupling coefficient.

Let us first study the Higgs potential of our model at zero temperature.
The tree-level Higgs potential of our model at zero temperature is essentially identical to the
tree-level Higgs potential of the MSSM.
However, the one-loop level Higgs potential of our model differs from the corresponding MSSM Higgs potential,
since additional contributions from the loops of
the vector-like quarks and the vector-like scalar quarks enter into
the radiative corrections to the Higgs potential of our model.
At zero temperature, the one-loop effective potential of our model may be written as
\begin{equation}
        V(\phi, 0) = V_0 (\phi, 0) + V_1 (\phi, 0) \ ,
\end{equation}
where $V_1(\phi, 0)$ is the radiative corrections at zero temperature.
It is given by the effective potential method as [30]
\begin{equation}
    V_1(\phi, 0) =  \sum_l {n_l  m_l^4 (\phi) \over 64 \pi^2}
            \left [ \log \left ({m_l^2 (\phi) \over \Lambda^2 } \right ) - {3 \over 2} \right ] \ ,
\end{equation}
where $l$ stands for the participating particles, namely,
the electroweak gauge bosons $W$ and $Z$, top quark $t$ and scalar top quarks
${\tilde t}_1$ and ${\tilde t}_2$,
as well as the vector-like quarks $T_1$ and $T_2$ and the vector-like scalar quarks
${\tilde T}_i$ ($i=1, 2, 3, 4$),
$\Lambda$ is the renormalization scale of the modified minimal subtraction scheme, set as $m_Z$,
and $n_l$ is the degree of freedom for the participating particles:
$n_W = 6$, $n_Z = 3$, $n_t = - 12$, $n_{{\tilde t}_i} = 6$ ($i=1, 2$),
$n_{T_1} = n_{T_2} = - 12$, $n_{{\tilde T}_i} = 6$ ($i=1, 2, 3, 4$).

Now, we calculate the tree-level masses of the relevant particles which will be used to
obtain the one-loop effective potential.
The masses of $W$ boson, $Z$ boson, and top quark, respectively, at the tree-level
are given as
\begin{eqnarray}
& & m_W^2(\phi) = {g_2^2 \over 2} \phi^2 \ , \quad
m_Z^2(\phi) = {(g_1^2 + g_2^2) \over 2 } \phi^2  \ , \quad
m_t^2(\phi) = h_t^2 \sin \beta^2 \phi^2   \ .
\end{eqnarray}

We assume that the lighter scalar top quark is predominantly right-handed.
As is well known, the scalar top quarks contribute to the tree-level Higgs sector considerably.
The masses of the scalar top quarks are given as [2,4],
\begin{eqnarray}
& & m_{{\tilde t}_1}^2 (\phi) = m_{U_3}^2 + D_R^2(\phi) + m_t^2 (\phi)
\bigg [1 - { {\tilde A}_t^2 \over m_{Q_3}^2} \bigg]  \ , \cr
& & \cr
& & m_{{\tilde t}_2}^2 (\phi) = m_{Q_3}^2 + D_L^2 (\phi) + m_t^2 (\phi)
\bigg [1 + { {\tilde A}_t^2 \over m_{Q_3}^2} \bigg]  \ ,
\end{eqnarray}
where $m_{Q_3}$ and $m_{U_3}$ are the SUSY breaking mass parameters,
$D_L^2 (\phi)$ and $D_R^2 (\phi)$ are the $D$-term contributions arising from
the SM gauge bosons, defined as
\begin{equation}
D_L^2 (\phi) = {3 g_2^2 - g_1^2 \over 12} \cos 2 \beta \phi^2 \ ,
\quad
D_R^2 (\phi) = {g_1^2 \over 3} \cos 2 \beta \phi^2 \ ,
\end{equation}
and ${\tilde A}_t = A_t - \mu \cot \beta$,
where $A_t$ is the trilinear SUSY breaking mass parameter.
Note that the difference in size between $D_L^2 (\phi)$ and $D_R^2 (\phi)$
causes  a relatively small mass splitting between ${\tilde t}_1$ and ${\tilde t}_1$,
whereas the opposite sign of ${\tilde A}_t$ in the above expressions indicates
that the mass eigenstates of scalar top quarks are mixtures of the weak eigenstates.

We also calculate the masses of the vector-like quarks and the vector-like scalar quarks at the tree level.
The mass matrix for the vector-like quarks at the tree level is given as $M_T M_T^{\dag}$ [28], with
\begin{equation}
M_T (\phi) =
\left(\begin{array}{cc}
M_Q & Y_u \sin \beta \phi  \\
Y_d \cos \beta \phi & M_U  \
\end{array} \right)  \ ,
\end{equation}
The eigenvalues of the mass matrix are the squared masses of the vector-like quarks.
If we take $M_Q = M_U$ for simplicity, they are given to a good approximation as:
\begin{eqnarray}
    && m_{T_1}^2(\phi) = M_Q^2 + Y_u^2 \sin^2 \beta \phi^2  \ ,  \cr
    && m_{T_2}^2(\phi) = M_Q^2 + Y_d^2 \cos^2 \beta \phi^2 \ .
\end{eqnarray}

The mass matrix for the vector-like scalar quarks are expressed as a $4\times 4$ matrix,
\begin{equation}
    M_S^2 = M_F^2 +
\left(\begin{array}{cccc}
m_Q^2 + D_L^2 (\phi) & 0 & b^2_Q & Y_u \sin \beta \phi {\tilde A}_u  \\
0 & m_U^2 - D_R^2 (\phi) & Y_d \cos \beta \phi {\tilde A}_d & b^2_U  \\
b^2_Q & Y_d \cos \beta \phi {\tilde A}_d &
m_{\bar Q}^2 - D_L^2 (\phi) - {{\displaystyle D_R^2 (\phi) } \over {\displaystyle 2}} & 0   \\
Y_u \sin \beta \phi {\tilde A}_u & b^2_U & 0 & m_{\bar U}^2 + D_R^2 (\phi)
\end{array} \right)  \ ,
\end{equation}
where
$b_Q$ and $b_U$ are the soft SUSY breaking mass,
$m_Q$, $m_{\bar Q}$, $m_U$ and $m_{\bar U}$ are the soft SUSY breaking masses,
${\tilde A}_u = A_u/Y_u - \mu \cot \beta$ and
${\tilde A}_d = A_d/Y_d - \mu \tan \beta$,
with $A_u$ and $A_d$ being the trilinear soft SUSY breaking mass parameters and
\begin{equation}
    M_F^2 =
\left(\begin{array}{cc}
M_T M_T^{\dag} & 0  \\
0 & M_T^{\dag} M_T   \
\end{array} \right)  \ .
\end{equation}
In subsequent calculations, we set $b_Q = b_U = 0$, $M_Q = M_U$, and adapt some approximations
for the mass matrices, following Ref. [28].

The squared masses of the four vector-like scalar quarks are given as the eigenvalues of their mass matrix.
We would like to reduce the number of free parameters, by absorbing $M_Q$ ($= M_U$) into
$m_Q$, $m_{\bar Q}$, $m_U$ and $m_{\bar U}$.
Thus, we may redefine $M_Q^2 + m_{\bar U}^2$ as $m_{\bar U}^2$,
$M_Q^2 + m_Q^2$ as $m_Q^2$, $M_Q^2 + m_U^2$ as $m_U^2$,
and $M_Q^2 + m_{\bar Q}^2$ as $m_{\bar Q}^2$.
Then, the squared masses of the vector-like scalar quarks are given in simplified expressions as
\begin{eqnarray}
& & m_{{\tilde T}_1}^2 (\phi) = m_{\bar U}^2 + D_R^2 (\phi) + Y_u^2 \sin^2 \beta \phi^2
\bigg [1 - { {\tilde A}_u^2 \over m_Q^2} \bigg]  \ , \cr
& & \cr
& & m_{{\tilde T}_2}^2 (\phi) = m_Q^2 + D_L^2 (\phi) + Y_u^2 \sin^2 \beta \phi^2
\bigg [1 + { {\tilde A}_u^2 \over m_Q^2} \bigg]  \ ,  \cr
& & \cr
& & m_{{\tilde T}_3}^2 (\phi) = m_U^2 - D_R^2 (\phi) + Y_d^2 \cos^2 \beta \phi^2
\bigg [1 - { {\tilde A}_d^2 \over m_{\bar Q}^2} \bigg]  \ , \cr
& & \cr
& & m_{{\tilde T}_4}^2 (\phi) = m_{\bar Q}^2 - D_L^2 (\phi)- {D_R^2 (\phi) \over 2} +
Y_d^2 \cos^2 \beta \phi^2
\bigg [1 + { {\tilde A}_d^2 \over m_{\bar Q}^2} \bigg]  \ .
\end{eqnarray}

We note that we keep the $D$-term contributions, both in the scalar top quark masses and in the
vector-like scalar quark masses.
Due to the presence of $D_L$ and $D_R$, these scalar quarks are not degenerate in mass in general.
If both $D_L = D_R = 0$ and ${\tilde A}_q = 0$ ($q = t,u,d$),
the scalar top quarks as well as the vector-like scalar quarks would be degenerate in mass.

We would like to note that  the strength of the electroweak phase transition tends to
decrease when the mixings between scalar quarks are taken into account.
Thus, for the first order EWPT to be strong, the mixings among the scalar quarks are not favored.
Nevertheless, we assume that ${\tilde A}_q$ ($q = t,u,d$) are not zero, implying that there are mixings
among scalar quarks.
Later, we will discuss on the effect of the mixings among the scalar quarks on the strength of the first-order EWPT.

The renormalized parameter $m_0^2$ in the Higgs potential can be expressed in terms of other parameters.
From the minimum condition of the first derivative of $V (\phi, 0)$ with respect to $\phi$,
$m_0^2$ at the one-loop level is expressed as
\begin{eqnarray}
    m_0^2
    & = & { 1 \over 2} m_Z^2 \cos^2 2 \beta
    + {3 m_W^4 \over 16 \pi^2 v^2 }  \left [ \log \left ({m_W^2 \over \Lambda^2} \right ) - 1 \right ]
    + {3 m_Z^4 \over 32 \pi^2 v^2 }  \left [\log \left ({m_Z^2 \over \Lambda^2} \right ) - 1 \right ] \cr
    & &\mbox{}
    - {3 m_t^4 \over 8 \pi^2 v^2 }  \left [ \log \left ({m_t^2 \over \Lambda^2} \right ) - 1 \right ]  \cr
    & &\mbox{}
    + {3 m_{{\tilde t}_1}^2 \over 16 \pi^2 v^2 } \bigg (m_{{\tilde t}_1}^2 - m_{U_3}^2 \bigg )
    \left [ \log \left ({m_{{\tilde t}_1}^2 \over \Lambda^2} \right ) - 1 \right ]
    + {3 m_{{\tilde t}_2}^2 \over 16 \pi^2 v^2 } \bigg (m_{{\tilde t}_2}^2 - m_{Q_3}^2 \bigg )
    \left [ \log \left ({m_{{\tilde t}_2}^2 \over \Lambda^2} \right ) - 1 \right ]  \cr
    & &\mbox{}
    - \sum_{l=1,2} {3 m_{T_l}^2 \over 8 \pi^2 v^2 } \bigg (m_{T_l}^2 - M_Q^2 \bigg )
    \left [ \log \left ({m_{T_l}^2 \over \Lambda^2} \right ) - 1 \right ] \cr
    & &\mbox{}
    + {3 m_{{\tilde T}_1}^2 \over 16 \pi^2 v^2 } \bigg (m_{{\tilde T}_1}^2 - m_{\bar U}^2 \bigg )
    \left [ \log \left ({m_{{\tilde T}_1}^2 \over \Lambda^2} \right ) - 1 \right ]   \cr
& &
    + {3 m_{{\tilde T}_2}^2 \over 16 \pi^2 v^2 } \bigg (m_{{\tilde T}_2}^2 - m_Q^2 \bigg )
    \left [ \log \left ({m_{{\tilde T}_2}^2 \over \Lambda^2} \right ) - 1 \right ] \cr
    & &\mbox{}
    + {3 m_{{\tilde T}_3}^2 \over 16 \pi^2 v^2 } \bigg (m_{{\tilde T}_3}^2 - m_U^2 \bigg )
    \left [ \log \left ({m_{{\tilde T}_3}^2 \over \Lambda^2} \right ) - 1 \right ]   \cr
& & + {3 m_{{\tilde T}_4}^2 \over 16 \pi^2 v^2 } \bigg (m_{{\tilde T}_4}^2 - m_{\bar Q}^2 \bigg )
    \left [ \log \left ({m_{{\tilde T}_4}^2 \over \Lambda^2} \right ) - 1 \right ]  \ ,
\end{eqnarray}
where the first term on the right-hand side is obtained from $V_0(\phi,0)$ and the remaining other terms are obtained from $V_1(\phi,0)$.
Also,  $v = \sqrt{v_d^2 + v_u^2} = 175$ GeV and
the field-dependant tree-level masses for the relevant particles are evaluated
at the electroweak symmetry breaking scale.
The mass of $\phi$ at the one-loop level is obtained by calculating the second derivative of $V (\phi, 0)$
with respect to $\phi$, evaluated at $\phi = v$, as
\begin{eqnarray}
m_{\phi}^2
    & = & m_Z^2 \cos^2 2 \beta
    + {3 m_W^4 \over 8 \pi^2 v^2} \log \left ( {m_W^2 \over \Lambda^2}\right )
    + {3 m_Z^4 \over 16 \pi^2 v^2} \log \left ( {m_Z^2 \over \Lambda^2}\right )
    - {3 m_t^4 \over 4 \pi^2 v^2} \log \left ( {m_t^2 \over \Lambda^2}\right )  \cr
    & &\mbox{}
    + {3 \over 8 \pi^2 v^2 } \bigg (m_{{\tilde t}_1}^2 - m_{U_3}^2 \bigg )^2
    \left [ \log \left ({m_{{\tilde t}_1}^2 \over \Lambda^2} \right ) - 1 \right ]  \cr
& &     + {3 \over 8 \pi^2 v^2 } \bigg (m_{{\tilde t}_2}^2 - m_{Q_3}^2 \bigg )^2
    \left [ \log \left ({m_{{\tilde t}_2}^2 \over \Lambda^2} \right ) - 1 \right ]  \cr
    & &\mbox{}
    - \sum_{l=1,2} {3 \over 4 \pi^2 v^2 } \bigg (m_{T_l}^2 - M_Q^2 \bigg )^2
    \left [ \log \left ({m_{T_l}^2 \over \Lambda^2} \right ) - 1 \right ]  \cr
    & &\mbox{}
    + {3 \over 8 \pi^2 v^2 } \bigg (m_{{\tilde T}_1}^2 - m_{\bar U}^2 \bigg )^2
    \left [ \log \left ({m_{{\tilde T}_1}^2 \over \Lambda^2} \right ) - 1 \right ]   \cr
& &
    + {3 \over 8 \pi^2 v^2 } \bigg (m_{{\tilde T}_2}^2 - m_Q^2 \bigg )^2
    \left [ \log \left ({m_{{\tilde T}_2}^2 \over \Lambda^2} \right ) - 1 \right ] \cr
    & &\mbox{}
    + {3 \over 8 \pi^2 v^2 } \bigg (m_{{\tilde T}_3}^2 - m_U^2 \bigg )^2
    \left [ \log \left ({m_{{\tilde T}_3}^2 \over \Lambda^2} \right ) - 1 \right ]   \cr
& & + {3 \over 8 \pi^2 v^2 } \bigg (m_{{\tilde T}_4}^2 - m_{\bar Q}^2 \bigg )^2
    \left [ \log \left ({m_{{\tilde T}_4}^2 \over \Lambda^2} \right ) - 1 \right ]  \ ,
\end{eqnarray}
where the first term on the right-hand side is the tree-level value and the remaining terms are
the radiative corrections.
The field-dependant tree-level masses for the relevant particles are evaluated
at the electroweak symmetry breaking scale, $v$.

Let us now study the Higgs potential of our model at finite temperature.
At temperature $T$, the full effective Higgs potential at the one-loop level may be written as
\begin{equation}
        V(\phi, T) = V_0(\phi,0) + V_1(\phi, 0) + V_1(\phi, T) \ ,
\end{equation}
where $V_1(\phi, T)$ is the the radiative correction at finite temperature.
Employing the effective potential method [31], we have
\begin{equation}
    V_1 (\phi, T)
    =  \sum_l {n_l T^4 \over 2 \pi^2}
            \int_0^{\infty} dx \ x^2 \
            \log \left [1 \pm \exp{\left ( - \sqrt {x^2 + {m_l^2 (\phi)/T^2 }} \right )  } \right ] \ ,
\end{equation}
where $l = W$, $Z$, $t$, ${\tilde t}_1$, ${\tilde t}_2$, $T_1$, $T_2$,
${\tilde T}_i$ ($i = 1,2,3,4$), and the negative sign is for bosons and the positive sign for fermions.
Note that the thermal effects due to $W$ boson, $Z$ boson, top quark,
vector-like quarks and their scalar partners are taken into account.
The possibly degenerate global minimum of $V(\phi, T)$ is defined as the vacuum at $T$.
The explicit expression for $V_1(\phi, T)$ is obtained by the exact numerical integration, instead of
the high-temperature approximation.

\section{Numerical Analysis}

It is easy to confirm that the above Higgs potential at the one-loop level becomes symmetric at very high temperature.
Also, one may check that the electroweak symmetry of the Higgs potential is broken at zero temperature.
In between the two extreme temperatures, there is an intermediate temperature,
called the critical temperature $T_c$,
at which the above Higgs potential has a doubly degenerate minimum.
Thus, at $T_c$, one minimum occurs at $\phi = 0$, which is the trivial minimum,
and the other minimum takes place at $\phi = v_c$, the critical VEV.
The trivial minimum represents the vacuum of the symmetric phase of the universe and
the minimum of $\phi = v_c$ represents the vacuum of the broken phase.
The EWPT may take place from the symmetric phase vacuum to the broken phase vacuum.

The critical temperature as well as the critical VEV is important for the successful EWPT,
since they determine the strength of the first-order EWPT.
Indeed, $v_c$ defined in this way might be a reasonable characterization of the strength of the phase transition.
The strength of the first-order EWPT, which is driven by a non-equilibrium (out-of-equilibrium) condition,
depends on the relative size between the critical temperature $T_c$
and the critical VEV $v_c$.
In general, the first-order EWPT is said to be strong if $v_c>T_c$.

For our numerical analysis, we fix the values of some parameters.
We set $\Lambda = m_Z$ for the renormalization scale
and $m_t = 175$ GeV for top quark mass.
Also, we take $m_{Q_3}^2 = 6 \times 10^5$ GeV$^2$, $m_{U_3}^2 = 10^4$ GeV$^2$,
${\tilde A}_t^2 = 6 \times 10^4$ GeV$^2$, which appear in the scalar top quark masses,
$M_Q^2 = 700^2$ GeV$^2$, $m_{\bar U}^2 = m_U^2 = 200^2$ GeV$^2$,
$m_Q^2 = m_{\bar Q}^2 = 2.4 \times 10^6$ GeV$^2$, and
${\tilde A}_u^2 = {\tilde A}_d^2 = 2.16 \times 10^6$ GeV$^2$,
which appear in the masses of the vector-like scalar quarks.

The Yukawa coupling coefficients of the vector-like quarks in our model
should satisfy the renormalization group (RG) equations
in order to avoid possible Landau poles at high energy scale, as discussed by Moroi and Okada [22,23].
The RG equations for $Y_u$ and $Y_d$ are solved numerically by using the Runge-Kutta method,
where no Landau pole is present.
The results for their evolutions are shown in Figs. 1a and 1b,
where, for simplicity, we take the same initial values for $Y_u$ and $Y_d$ at the electroweak scale:
$Y_u = Y_d = 0.87$, 0.9, 1.0, 1.1, and 1.13.

Notice that if the initial values for $Y_u$ and $Y_d$ are not larger than 1.0,
both $Y_u$ and $Y_d$ become smaller for higher energy scale.
However, if they start from initial values larger than 1.0, as Figs. 1a and 1b show,
they tend to increase as the energy scale increases.
This is because their renormalization beta functions are dominated by the positive
contributions from themselves which are larger than the negative contributions from the gauge
couplings.

With these values, we calculate the value of $V(\phi, T)$ as a function of $\phi$
at given temperature $T$, such that $V(0,T_c) = V(v_c, T_c) = 0$ and
\[
 {\partial V \over \partial \phi}|_{\phi = 0}
= {\partial V \over \partial \phi}|_{\phi = v_c} =0 \ .
\]
These two conditions establish the doubly degenerate vacua of the symmetric phase and the broken phase.

In this way, we obtain Fig. 2 for $\tan \beta = 5$ and $Y_u = Y_d = 1$, at $T = 157.58$ GeV,
where $V(\phi, T)$ shows the typical shape for the first-order EWPT.
One can see that there are two minima for $V(\phi, T)$, at around $\phi = 0$ and $\phi = 258$ GeV.
Since the values of $V(\phi, T)$ at these two minima are equal,
they are doubly degenerate vacua with a potential barrier between them.
The first-order EWPT may take place from one vacuum to the other one through thermal tunneling.
Therefore, $\phi = 258$ GeV is the critical VEV for the vacuum of the broken phase and
$T = 157.58$ GeV is the critical temperature $T_c$
that allows the phase transition from the symmetric-phase state to the broken-phase state.
From these values, we obtain $v_c/T_c \sim 1.637$, which certainly implies that
the first-order EWPT shown in Fig. 2 is strong.
Consequently, Fig. 2 suggests that there is at least one set of parameter values of our model
at the one-loop level which allows the possibility of a strongly first-order EWPT.

For the set of parameter values of Fig. 2, the masses of participating particles are evaluated.
The mass of the lightest scalar Higgs boson is calculated to be $m_{\phi} = 151.41$ GeV,
whereas the masses of the other Higgs bosons are all very heavy in the decoupling limit.
The masses of the scalar top quarks are obtained as $m_{{\tilde t}_1} = 190.67$ GeV
and $m_{{\tilde t}_2} = 794.35$ GeV.
Note that both of two scalar top quarks are heavier than top quark.
In other words, our model does not require a light scalar top quark in order to realize
a strongly first-order EWPT.

Vector-like quarks and scalar quarks are characteristic members of our model.
Their masses are calculated to be $m_{T_1} = 720.72$ GeV and $m_{T_2} = 700.84$ GeV.
The masses of the vector-like scalar quarks are $m_{{\tilde T}_1} = 200.62$ GeV,
$m_{{\tilde T}_2} = 1566.76$, $m_{{\tilde T}_3} = 203.28$, and $m_{{\tilde T}_4} = 1550.97$.
These masses might affect the precision electroweak observables, due to the virtual corrections at the
one-loop level.
As noticed by Ref. [28], the most important contributions from the vector-like quark sector to
the precision electroweak observables may be described in terms of two parameters, $S$ and $T$,
which have been proposed by Peskin and Takeuchi [32].
We follow the method of Ref. [28] to calculate the contributions from the vector-like quark sector to
the electroweak gauge boson self energies.
We find that the corrections are practically negligible:
$\Delta S = 0.00187$ and $\Delta T = 0.00686$ for the parameter values in Fig. 2.
Thus, the masses of the vector-like quark sector obtained from the parameter values in Fig. 2 are
consistent with the electroweak precision test.

Now, we would like to show that the above set of parameter values is not singular.
We examine other regions of the parameter space to find that a large number of sets of
parameter values within the parameter space allow the strongly first-order EWPT.
As an illustration, we vary the value of $\tan\beta$, while fixing the values of other parameters,
and repeat the above procedure.
Our result is listed in Table 1, where one may note that a wide range of $\tan\beta$
indeed allow the desired phase transition.

\begin{table}[ht]
\caption{The critical temperature and the critical VEV for the strongly first-order EWPT
for some values of $1 < \tan \beta < 40$.
The values of the other parameters are the same as Fig. 2,
in particular, ${\tilde A}_u^2 = {\tilde A}_d^2 = 2.16 \times 10^6$ GeV$^2$  and $Y_u = Y_d = 1$.}
\begin{center}
    \begin{tabular}{c||c|c|c|c|c|c|c}
    \hline
    \hline
    $\tan \beta$ &  1 & 1.5 & 2 & 5 & 10  & 30 & 40 \\
    \hline
    \hline
    $T_c$ (GeV) & 146.19 & 150.83 & 153.34 & 157.58 & 158.43 & 158.69 & 158.71 \\
    \hline
    $v_c$ (GeV) & 289 & 285 & 280 & 258 & 255 & 254 & 253 \\
    \hline
    $m_{\phi}$ (GeV) & 107.84 & 116.27 & 127.81 & 151.41 & 156.15 & 157.64 & 157.73 \\
    \hline
    $v_c/T_c$ & 1.9768 & 1.8895 & 1.8260 & 1.6372 & 1.6095 & 1.6006 & 1.5941\\
    \hline
    $m_{{\tilde t}_1}$ (GeV) & 193.81 & 192.51 & 191.77 & 190.67 & 190.48 & 190.41& 190.41 \\
    \hline
    $m_{{\tilde t}_2}$ (GeV) & 796.04 & 795.33 & 794.94 & 794.35 & 794.24 & 794.21 & 794.21\\
    \hline
    $m_{T_1}$ (GeV) & 710.85 & 714.98 & 717.28 & 720.72 & 721.33 & 721.51 & 721.53 \\
    \hline
    $m_{T_2}$ (GeV) & 700.85 & 706.69 & 704.36 & 700.84 & 700.21 & 700.02 & 700.01\\
    \hline
    $m_{{\tilde T}_1}$ (GeV) & 203.79 & 202.47 & 201.73 & 200.62 & 200.42 & 200.36 & 200.36 \\
    \hline
    $m_{{\tilde T}_2}$ (GeV) & 1558.55 & 1561.97 & 1563.89 & 1566.76 & 1567.26 & 1567.42 & 1567.43 \\
    \hline
    $m_{{\tilde T}_3}$ (GeV) & 203.79 & 203.57 & 203.45 & 203.28 & 203.24 & 203.23 & 203.23 \\
    \hline
    $m_{{\tilde T}_4}$ (GeV) & 1558.55 & 1555.40 & 1553.63 & 1550.97 & 1550.50 & 1550.36 & 1550.35 \\
    \hline
    \hline
    \end{tabular}
\end{center}
\end{table}

After establishing the possibility of the strongly first-order EWPT,
we are now interested in the behavior of the strength of the first-order EWPT with respect to
the mass of the lightest scalar Higgs boson.
It has been observed in both the SM and the MSSM that the strength of the first-order EWPT
always increases as the lightest scalar Higgs boson mass decreases.
Also, in the non-MSSMs with a Higgs singlet, a stronger first-order EWPT
is allowed for a lighter $S_1$, where $S_1$ is the lightest scalar Higgs boson [6-19].
Thus, one can expect that in our model too the strength of the EWPT is enhanced as
the lightest Higgs boson becomes light.

In Table 1, one can note that $m_{\phi}$ increases as $\tan\beta$ increases from 1 to 40.
At the same time, $v_c/T_c$ decreases as $\tan\beta$ increases from 1 to 40.
Thus, for the set of parameter values in Table 1, one may notice the tendency of
stronger phase transition for smaller $m_{\phi}$.
In other words, the behavior of our model is consistent those of the SM, the MSSM, or the non-MSSMs
in the sense that the smaller mass of the lightest Higgs boson may indeed enhance
the strength of the first-order EWPT.

Another point we would like to discuss is the role of the Higgs cubic term that appears in the power expansion
of the one-loop Higgs potential at finite temperature, in the high-temperature approximation.
The size of this Higgs cubic term is essential to the strength of the first-order EWPT.
The coefficient of this term depends on a number of bosons such as the SM gauge bosons, the scalar top
quarks, and the vector-like scalar quarks.
The SM gauge bosons may take part in the coefficient of the $\phi^3$ term in the SM.
However, in the MSSM, its contribution to the coefficient of the $\phi^3$ term is not so much important
as the contributions from the scalar top quarks, as long as the scalar top quarks are sufficiently light.

If the masses of the scalar top quarks are very large (1 or 2 TeV) in the MSSM, its contribution to the strength of
the first-order EWPT through the Higgs cubic term would be negligibly small.
On the other hand, if they are sufficiently light at the electroweak scale
(for example, about 165 GeV between the top quark mass
and the experimental lower bound of about 155 GeV), it might significantly contribute to the strength of the
first-order EWPT.
Thus, in the MSSM, at least one of the scalar top quarks need to be lighter than top quark in order to
achieve a strongly first-order EWPT.

The strength of the first-order EWPT in our model depends also on the vector-like scalar quarks.
As shown in Eq. 15, the pattern of the masses of the vector-like scalar quarks may be represented by
a sum of the soft SUSY mass and the Yukawa term, up to a constant.
The $D^2$ terms are constant if the gauge coupling coefficients remain as constant.

If the Yukawa coupling coefficients are small ($ <0.1$) and the soft SUSY masses are large ($> 1$ TeV),
the masses of the vector-like scalar quarks would be large ($> 1$ TeV),
and the loop contributions of the vector-like scalar quarks to the Higgs cubic term
would be very small.
Thus, in this case, the vector-like scalar quarks suppress the strength of the first-order EWPT,
by suppressing the coefficient of the $\phi^3$ term.
In fact, in the symmetric phase state where $\phi$ becomes zero, the vector-like scalar quarks
do not become massless but remain heavy at TeV scale.
In other words, large soft SUSY masses prevent the appearance of the Higgs cubic term in the
expansion of the effective thermal potential.

On the other hand, if the Yukawa coupling coefficients are large ($ \sim 1$) and the soft SUSY
masses are small ($\sim 200$ GeV),
some of the vector-like scalar quarks would be light ($ \sim 200$ GeV),
In this case, the loop contributions of the vector-like scalar quarks to the Higgs cubic term
would be large, and thus enhance the strength of the first-order EWPT.
Actually, our choice of parameter values are large Yukawa coupling coefficients and small soft SUSY masses
such that a strongly first-order EWPT is possible.

The situation is quite similar in our model, where there are additional scalar fields,
namely, vector-like scalar quarks.
When $V_1(\phi,T)$ in our model is expanded in powers of $\phi$ in the high-temperature approximation,
there is a term proportional to $\phi^3$, arising from the vector-like scalar quarks
as well as from the SM gauge bosons and scalar top quarks.
The additional contribution due to loops of vector-like scalar quarks
certainly enhance the strength of the first-order EWPT,
thus relieving the necessity of a light scalar top quark in our model.

Next, let us study the effect of the mixing in the scalar quarks on the strength of the first-order EWPT.
The mass eigenstates of scalar quarks are given as mixtures of the weak eigenstates,
since the mass matrix for them is not diagonal in general.
The size of the mixing between them is determined by the size of the off-diagonal matrix elements.
In our model, the mixing between the scalar top quarks is determined by ${\tilde A}_t^2$
and the mixing among the vector-like scalar quarks is determined by ${\tilde A}_u^2 = {\tilde A}_d^2$.
We show in Table 2 the result of varying these parameters, while keeping other parameters fixed.

\begin{table}[ht]
\caption{The critical temperature and the critical VEV for the strongly first-order EWPT
for two different values of ${\tilde A}_u^2 = {\tilde A}_d^2$.
The values of the other parameters are the same as Table 1, in particular, $\tan\beta =5$.
Thus, the first column is identical to the fourth column of Table 1.}
\begin{center}
    \begin{tabular}{c||c|c}
    \hline
    \hline
    ${\tilde A}_u^2 = {\tilde A}_d^2$ (GeV$^2$) &  $2.16 \times 10^6$ & $1.2 \times 10^6$ \\
    \hline
    \hline
    $T_c$ (GeV) & 157.58 & 177.65 \\
    \hline
    $v_c$ (GeV) & 258 & 522 \\
    \hline
    $m_{\phi}$ (GeV) & 151.41 & 122.84 \\
    \hline
    $v_c/T_c$ &1.6372 & 2.9383 \\
    \hline
    $m_{{\tilde t}_1}$ (GeV) & 190.67 & 190.67 \\
    \hline
    $m_{{\tilde t}_2}$ (GeV) & 794.35 & 794.35 \\
    \hline
    $m_{T_1}$ (GeV) & 720.72 & 720.72 \\
    \hline
    $m_{T_2}$ (GeV) & 700.84 & 700.84 \\
    \hline
    $m_{{\tilde T}_1}$ (GeV) & 200.62 & 228.09 \\
    \hline
    $m_{{\tilde T}_2}$ (GeV) & 1566.76 & 1562.99 \\
    \hline
    $m_{{\tilde T}_3}$ (GeV) & 203.28 & 203.28 \\
    \hline
    $m_{{\tilde T}_4}$ (GeV) & 1550.97 & 1550.82 \\
    \hline
    \hline
    \end{tabular}
\end{center}
\end{table}
The result in Table 2 suggests that the first-order EWPT becomes much stronger
if the value of the parameter for the mixing among vector-like scalar quarks is small.

\section{Conclusions}

By introducing the vector-like quarks and scalar quarks to the MSSM,
we study the possibility of a strongly first-order EWPT, at the one-loop level.
The Higgs sector of our model is essentially identical to the MSSM Higgs sector,
while there are additionally two vector-like quarks and four vector-like scalar quarks in our model.
At the one-loop level, we take into account the thermal effects due to loops of $W$ boson,
$Z$ boson, top quark, scalar top quarks, vector-like quarks, and vector-like scalar quarks.

We find that a number of parameter sets in the reasonably wide parameter space of our model allow
the strongly first-order EWPT, thus satisfying one of the Sakharov conditions for electroweak baryogenesis.
The strength of the first-order EWPT is very important for the successful electroweak baryogenesis.
In the MSSM, a strongly first-order EWPT is guaranteed by a light scalar top quark; whose mass
should be smaller than the top quark mass.
In other words, it is experimentally challenging for the MSSM to discover a scalar top quark whose mass
is smaller than the top quark mass, if the first-order EWPT should be strong enough.
On the other hand, in our model, the masses of the scalar top quarks need not be smaller than top quark.

We work in the limit where all the Higgs bosons except for the lightest scalar Higgs boson
are as heavy as the SUSY scale of a few TeV.
In this limit, the lightest scalar Higgs boson behaves just like the SM Higgs boson whereas the other
Higgs bosons are decoupled from the phenomenology at the electroweak scale.
We note that, in the MSSM, the strength of the first-order EWPT increases as
the mass of the pseudoscalar Higgs boson increases.
Thus, the decoupling limit may be considered as an optimal situation for the first-order EWPT to be sufficiently strong,
since the mass of the pseudoscalar Higgs boson is already very large.
However, the decoupling limit alone may not be enough for the MSSM to achieve the desired phase transition,
since in this limit the MSSM Higgs sector resembles the SM Higgs sector and it has already been observed
that the SM Higgs sector cannot successfully generate the baryon asymmetry of the universe.
Hence, the MSSM requires a different mechanism in addition to the decoupling limit
for the strongly first-order EWPT.
Our study suggests that the introduction of vector-like quark sector may be a possible mechanism.
The existence of the vector-like quarks and scalar quarks not only enhance the strength of
the first-order EWPT in our model, but also alleviate the requirement that the mass of a scalar top quark
should be lighter than the top quark mass.

The strength of the first-order EWPT in our model is affected by the values of the relevant parameters.
Our numerical analysis shows that it is enhanced as the mass of the lightest scalar Higgs boson decreases.
Also, if we reduce the value of ${\tilde A}_d = {\tilde A}_u$,
the parameters that account for the mixing among the vector-like scalar quarks,
the strength of the EWPT increases.

\section*{Acknowledgments}

S. W. Ham thanks J. Y. Lee, P. Ko, and Y. G. Kim for valuable comments.
He would like to acknowledge the support from KISTI under
"The Strategic Supercomputing Support Program (No. KSC-2008-S01-0011)"
with Dr. Kihyeon Cho as the technical supporter.
This research was supported by Basic Science Research Program
through the National Research Foundation of Korea (NRF) funded
by the Ministry of Education, Science and Technology (2009-0086961).



\vfil\eject

\renewcommand\thefigure{1a}
\begin{figure}[t]
\begin{center}
\includegraphics[scale=0.6]{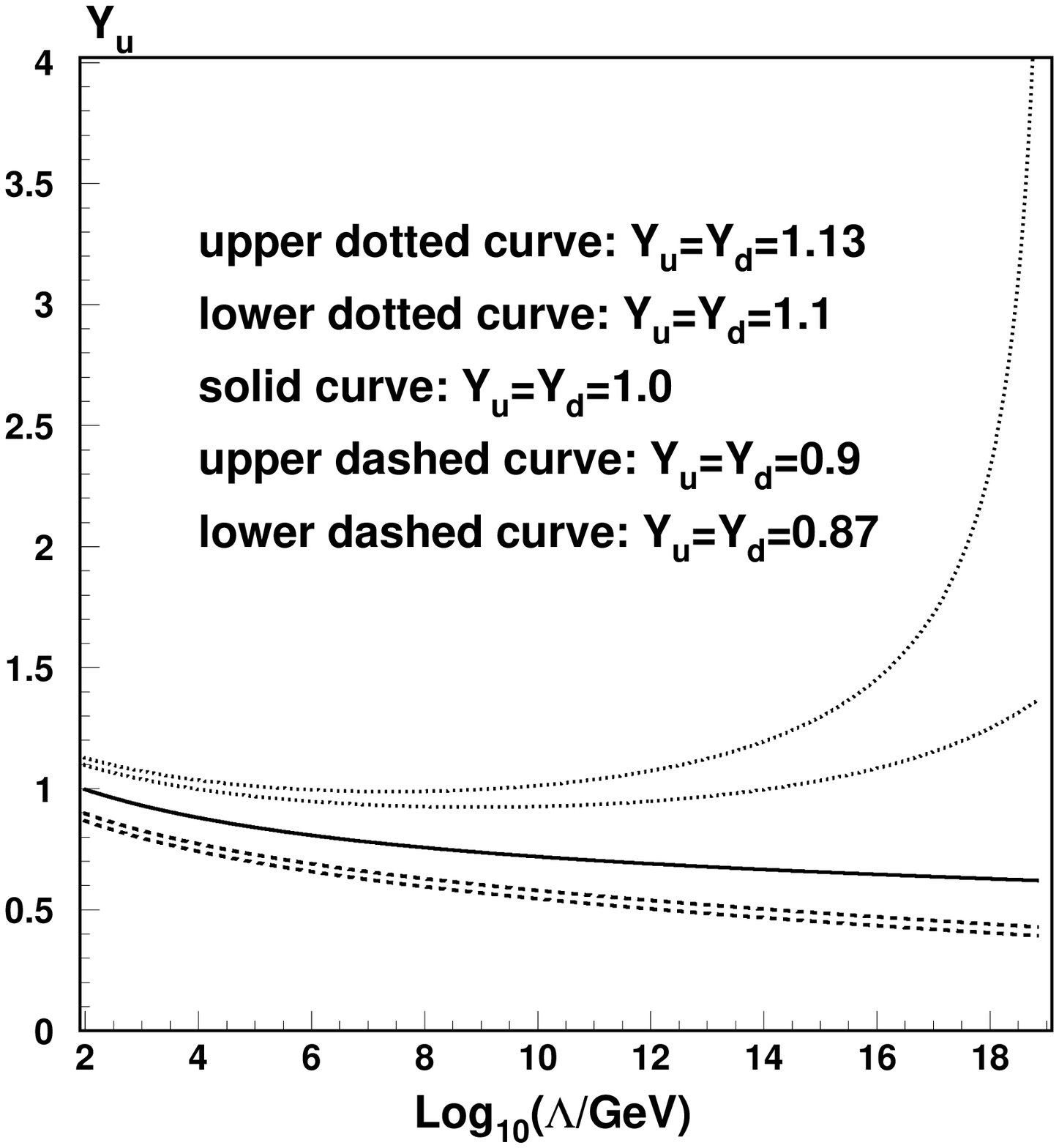}
\caption[plot]{The RG evolutions for $Y_u$ by using Runge-Kutta
method for different initial values of $Y_u$ and $Y_d$ at the
electroweak scale. }
\end{center}
\end{figure}

\renewcommand\thefigure{1b}
\begin{figure}[t]
\begin{center}
\includegraphics[scale=0.6]{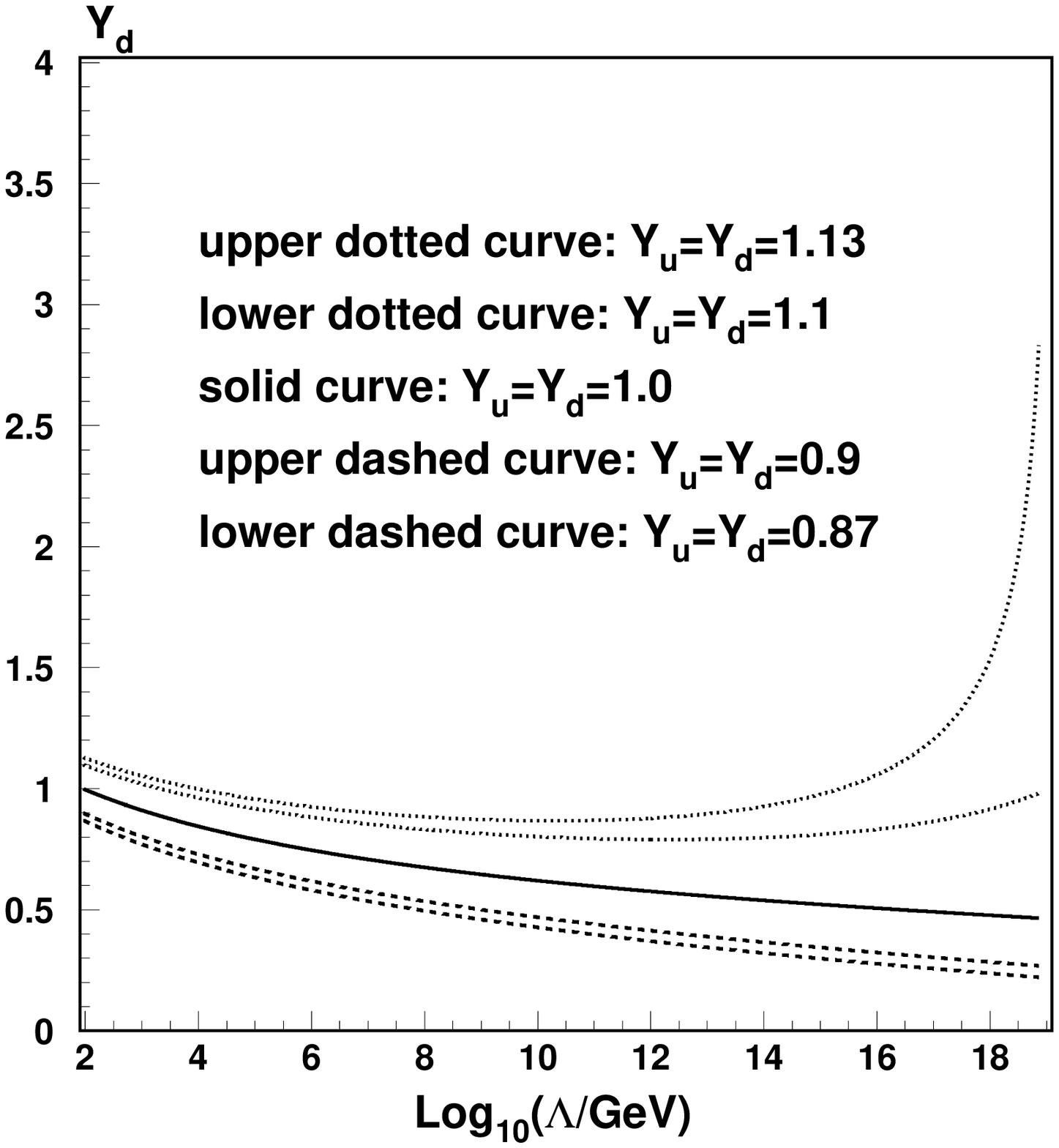}
\caption[plot]{The RG evolutions for $Y_d$ by using Runge-Kutta
method for different initial values of $Y_u$ and $Y_d$ at the
electroweak scale.}
\end{center}
\end{figure}

\renewcommand\thefigure{2}
\begin{figure}[t]
\begin{center}
\includegraphics[scale=0.6]{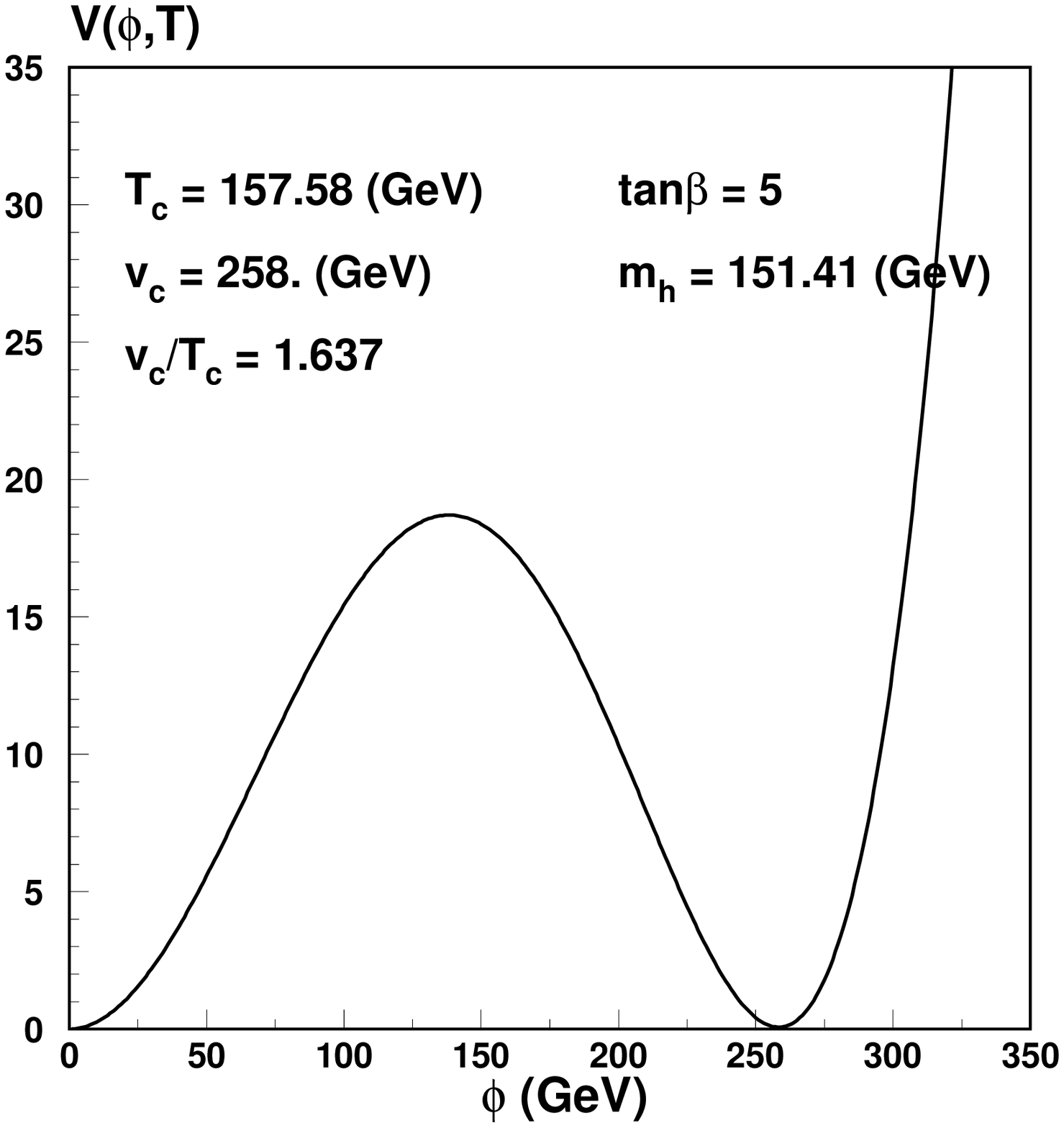}
\caption[plot]{The values of the relevant parameters are
$m_{Q_3}^2 = 6 \times 10^5$ GeV$^2$, $m_{U_3}^2 = 10^4$ GeV$^2$,
${\tilde A}_t^2 = 6 \times 10^4$ GeV$^2$,
$M_Q^2 = 700^2$ GeV$^2$, $m_{\bar U}^2 = m_U^2 = 200^2$ GeV$^2$,
$m_Q^2 = m_{\bar Q}^2 = 2.4 \times 10^6$ GeV$^2$, and
${\tilde A}_u^2 = {\tilde A}_d^2 = 2.16 \times 10^6$ GeV$^2$,
and $Y_u = Y_d = 1.0$ at the electrowak scale.}
\end{center}
\end{figure}

\end{document}